\documentstyle[epsf]{article}
\begin{document}
\title{Monopole-Antimonopole Solutions
of the Skyrmed $SU(2)$ Yang-Mills-Higgs Model}
\author{{\large Vanush Paturyan}$^{\star\dagger}$
and {\large D. H. Tchrakian}$^{\dagger\ddagger}$ \\ \\
\medskip
$^{\star}${\small Department of Computer Science,
National University of Ireland Maynooth,}\\
{\small Maynooth, Ireland}\\ \\
$^{\dagger}${\small Department of
Mathematical Physics, National University of Ireland Maynooth,} \\
{\small Maynooth, Ireland}\\ \\
$^{\ddagger}${\small School of Theoretical Physics, Dublin Institute for
Advanced Studies}\\
{\small 10 Burlington Road, Dublin 4, Ireland}}

\date{\today}
\newcommand{\dd}{\mbox{d}}\newcommand{\tr}{\mbox{tr}}
\newcommand{\ee}{\end{equation}}
\newcommand{\be}{\begin{equation}}
\newcommand{\ii}{\mbox{i}}\newcommand{\e}{\mbox{e}}
\newcommand{\pa}{\partial}\newcommand{\Om}{\Omega}
\newcommand{\bfph}{{\bf \phi}}
\newcommand{\lm}{\lambda}
\newcommand{\st}{\sin\theta}
\newcommand{\ct}{\cos\theta}
\newcommand{\snt}{\sin 2\theta}
\newcommand{\cnt}{\cos 2\theta}

\def\theequation{\arabic{equation}}
\renewcommand{\thefootnote}{\fnsymbol{footnote}}
\newcommand{\re}[1]{(\ref{#1})}
\newcommand{\bfR}{{\sf R\hspace*{-0.9ex}\rule{0.15ex}%
{1.5ex}\hspace*{0.9ex}}}
\newcommand{\N}{{\sf N\hspace*{-1.0ex}\rule{0.15ex}%
{1.3ex}\hspace*{1.0ex}}}
\newcommand{\Q}{{\sf Q\hspace*{-1.1ex}\rule{0.15ex}%
{1.5ex}\hspace*{1.1ex}}}
\newcommand{\C}{{\sf C\hspace*{-0.9ex}\rule{0.15ex}%
{1.3ex}\hspace*{0.9ex}}}
\renewcommand{\thefootnote}{\arabic{footnote}}
\renewcommand{\textfraction}{0.9}

\maketitle
\begin{abstract}

Axially symmetric
monopole anti-monopole dipole solutions to the second order equations of a
simple SU(2) Yang-Mills-Higgs model featuring a quartic Skyrme-like term are
constructed numerically. The effect of varying the Skyrme coupling
constant on these solutions is studied in some detail.

\end{abstract}
 \vfill
 \vfill\eject

\section{Introduction}

The SU(2) Georgi-Glashow model in the Bogomol'nyi-Prasad-Sommerfield (BPS)
limit supports monopoles~\cite{tHooft,Poly} which are solutions of the
first order self-duality equations~\cite{PS,Bog}. Away from the BPS
limit, when new gauge invariant and positive definite terms are added,
the resulting monopoles are described by the solutions to the second
order Euler--Lagrange equations, and not to the first order self-duality
equations. Once these terms are introduced to the model, the BPS
topological bound cannot be saturated.

BPS and non-BPS monopoles differ in two remarkable respects. First, the
BPS multimonopoles can be constructed
analytically~\cite{RR,Ward,Forg,Pras} while the non-BPS monopoles, e.g.
when the Higgs potential is present~\cite{tHooft,Poly}, can only be
constructed numerically. Secondly, and perhaps physically more
interestingly, BPS monopoles do not interact while non-BPS monopoles
interact. In the presence of a Higgs potential this interaction is known
to be repulsive~\cite{M,N} and has been verified to be so
numerically~\cite{KKT}, while in the presence of Skyrme like terms,
higher order in both the Yang-Mills (YM) curvature and the Higgs covariant
derivatives, this interaction can be both repulsive and 
attractive~\cite{KOT}. In a particularly simple such (Skyrme like)
model, this interaction was found~\cite{GST} to be strictly attractive,
and moreover it was found~\cite{GST}, rather unexpectedly, that the
lowest energy bound states were the axially symmetric ones and not those
with Platonic symmetries. (It was unexpected since this feature
contrasts with that for Skyrmion bound states~\cite{BS}.)

All the above monopole solutions discussed are stable relative to the
topological lower bound whether they saturate this bound, as for the
BPS monopoles, or not, as for non-BPS ones. There is however another
class of non--selfdual solutions to the second order Euler--Lagrange
equations which are not stable and represent states of monopoles and
anti-monopoles in equilibrium. The existence of such solutions was first
proved by Taubes~\cite{Taubes} for the model featuring no Higgs potential (and
of course no higher order terms in the curvature and covariant derivative),
namely for the model which supports BPS multimonopoles. Such a non-BPS
solution, namely an unstable solution of the second order equations, was first
constructed for this system with $SU(3)$ gauge group and subject to spherical
symmetry by Burzlaff~\cite{B}. More recently Ioannidou and
Sutcliffe~\cite{IS} emoplyed a harmonic map Anstaz to construct such
spherically symmetric solutions to the same (BPS) system with gauge groups
$SU(3)$, $SU(4)$ and $SU(N)$. Using results on sigma model instantons, these
authors~\cite{IS} also argued that the zero charge solutions they constructed
described monopole anti-imonopole pairs.

A direct approach to constructing zero charge monopole anti-monopole pairs
for the $SU(2)$ BPS model was used sometime ago by R\"uber~\cite{Rueb}.
This was the numerical construction of axially symmetric solutions with
suitable boundary value conditions. More recently Kleihaus and
Kunz~\cite{KK} constructed this zero charge solution
for the full Georgi-Glashow model featuring a Higgs potential, and they
studied the effect of the Higgs potential in detail.
To date, no such study has been reported in the
literature pertaining to the model featuring higher order Skyrme like
terms. In the background of the above described scenario it is
pertinent to carry out such a study.

This is the aim of the present work. We will consider the zero charge
axially symmetric monopole anti-monopole solutions as in \cite{Rueb,KK},
for the simple skyrmed Higgs model studied in \cite{GST} whose axially
symmetric charge-2 monopoles are mutually attractive. This contrasts with the
monopole anti-monopole solutions studied in \cite{KK} for the model
whose charge-2 monopoles are mutually repulsive, which
makes the comparison of our results with those of \cite{KK} interesting.
In addition to constructing the vorticity-1 monopole anti-monopole
solutions, as in \cite{Rueb} and \cite{KK}, but now for the skyrmed model here,
we also construct the corresponding vorticity-2 solutions.

\boldmath
\section{Skyrmed $SU(2)$ Yang-Mills-Higgs Model}
\unboldmath

The static energy of the simplified Skyrme like model considered is
\begin{eqnarray}
{\cal E} & = &
\int\left\{
 \frac{1}{2}{\rm Tr}\{F_{\mu\nu}F^{\mu\nu}\}
+\frac{1}{4}{\rm Tr}\{D_\mu\Phi D^\mu\Phi\}
+\frac{\kappa}{8}
{\rm Tr}\{[D_\mu\Phi, D_\nu\Phi][D^\mu\Phi, D^\nu\Phi]\}
+\frac{\lambda}{2}{\rm Tr}\{(\Phi^2-\eta^2)^2\}
\right\} d^3r
\label{lag1}
\  \end{eqnarray}
with field strength tensor of the $su(2)$ gauge potential 
$A_\mu = \frac{1}{2} \tau_a A_\mu^a$,
\begin{eqnarray}
F_{\mu\nu} & = & \partial_\mu A_\nu - \partial_\nu A_\mu 
                  + i g\left[A_\mu, A_\nu \right] \ ,
\label{Fdef}
\end{eqnarray}
and covariant derivative
of the Higgs field $\Phi = \tau_a \phi^a$ in the adjoint representation
\begin{eqnarray}
D_\mu \Phi & = & \partial_\mu \Phi + i g\left[ A_\mu, \Phi \right]
\label{DPdef} \ , \end{eqnarray}
and $g$ denotes the gauge coupling constant, $\kappa$ the coupling
strength of the quartic Skyrme like Higgs kinetic term,
$\lambda$ the strength of the Higgs potential and
$\eta$ the vacuum expectation value of the Higgs field.

The topological charge $Q$ is the well known quantity
\begin{equation}
Q = \frac{1}{4\pi\eta}\varepsilon^{ijk}
\int {\rm Tr}\left\{F_{ij} D_k\Phi\right\} d^3r
\ , \label{Q} \end{equation}
corresponding to the magnetic charge $m=Q/g$, and takes integer
values that equal the winding number of the Higgs field~\cite{AFG}. The
latter is encoded with the boundary conditions which yield the value of
this integer.

To construct axially symmetric solutions that describe systems of monopoles and
multimonopoles, specific boundary conditions must be imposed the Higgs field
at infinity. For usual multimonopoles, the Higgs field at infinity is
described by the vortex number $n$ winding the azimuthal angle $\varphi$,
$n$ times and the polar angle $\theta$ does not wind. Zero magnetic charge
monopoles on the other hand, namely those we seek to construct, can be
achieved by requiring that in the asymptotic Higgs field the polar angle
is enhanced by another integer $m$. This can also be achieved automatically
by incorporating this integer $m$ in the Ansatz~\cite{Rueb,KK} as will be done
below. The integral \re{Q} can be evaluated for a system with $m$ zeros of the
Higgs field (i.e. with $m$ monopole and antimonopole centres), and with
vorticity $n$, yielding
\be
\label{mn}
Q=4\pi n\eta^3[1-(-1)^m]\ .
\ee
In this paper we will restrict to the charge zero case $m=2$ with
vorticity $n=1$, to carry out our detailed analysis of the system,
with special attention to the $\kappa$ dependence of the solutions.
After that, we will briefly study also the case of $n=2$ vorticity,
again with $m=2$. These are both monopole anti-monopole solutions to
the second order equations carrying $Q=0$.

\section{Static axially symmetric $Q=0$ Ansatz}

We choose the static, axially symmetric, purely magnetic Ansatz 
employed in \cite{Rueb} for the monopole-antimonopole solution
and in \cite{Klink,KuBri} for the sphaleron-antisphaleron
solution of the Weinberg-Salam model.
Here the gauge field is parametrized by
\begin{equation}
A_0 = 0\ , 
\ \ A_r =\frac{H_1}{2gr}\tau_\varphi^{(n)}\ , 
\ \ A_\theta = \frac{\left(1-H_2\right)}{g} \tau_\varphi^{(n)}, 
\ \ A_\varphi = -n\frac{\sin\theta}{g}\left(H_3 \tau^{(2,n)}_r+
\left(1-H_4\right)\tau^{(2,n)}_\theta\right)\  \ ,
\label{A_an}
\end{equation}
and the Higgs field by
\begin{equation}
\Phi =\eta
\left( \Phi_1 \tau^{(2,n)}_r \ + \Phi_2 \tau^{(2,n)}_\theta\right) \ .
\label{Phi_an}
\end{equation}
All functions $H_1,H_2,H_3,H_4,\Phi_1$ and $\Phi_2$ depend on $(r,\theta)$
or equivalently on $(\rho=r\sin\theta,z=r\cos\theta)$, with the $su(2)$
matrices $\tau^{(2,n)}_r$, $\tau^{(2,n)}_\theta$ and $\tau_\varphi^{(n)}$
defined in terms of the Pauli matrices $\tau_1,\tau_2,\tau_3$ as
\begin{eqnarray}
\tau^{(2,n)}_r & = & \sin 2\theta(\cos n\varphi\, \tau_1 + \sin n\varphi\,
\tau_2) + \cos 2\theta\, \tau_3 \ ,
\nonumber\\             
\tau^{(2,n)}_\theta & = & \cos 2\theta(\cos n\varphi\, \tau_1
+ \sin n\varphi\,\tau_2)- \sin 2\theta\, \tau_3 \ ,
\nonumber\\             
\tau_\varphi^{(n)} & = & -\sin n\varphi\, \tau_1 + \cos n\varphi\, \tau_2 \ ,
\label{n}
\end{eqnarray}
and for later convenience we define 
\begin{equation}
\tau^{(n)}_\rho  =  \cos n\varphi\, \tau_1 + \sin n\varphi\, \tau_2 \ .
\nonumber
\end{equation}
Note that the dependence on the vorticity $n$ is encoded through
$\tau^{(2,n)}_r$ and $\tau^{(2,n)}_\theta$, and of course $\tau^{(n)}_\rho$.

We change to dimensionless coordinates, Higgs field and coupling
parameters
by rescaling
$$
r \rightarrow \frac{r}{g \eta} \ , \ \ \
\Phi\rightarrow \eta \Phi \ , \ \ \
\kappa \rightarrow \frac{\kappa}{g^2\eta^4} \ , \ \ \
\lambda\rightarrow \frac{\lambda}{g^2} \  ,
$$
respectively.
Then this Ansatz leads to the field strength tensor 
\begin{eqnarray}
F_{r\theta} & = & 
 -\frac{1}{2 r}\left( \partial_\theta H_1 + 2 r \partial_r H_2\right)
                   \tau^{(n)}_\varphi \ ,
\nonumber\\
F_{r\varphi}  & = &  \frac{n}{2 r}\left\{
            \left( \snt H_1 - 2\st  H_1 (1-H_4) -2 \st r \partial_r H_3\right)
            \tau^{(2,n)}_r
            \right.
\nonumber\\ & & 
            \left.
           + \left( \cnt H_1 + 2\st  H_1 H_3     +2 \st r \partial_r H_4\right)
           \tau^{(2,n)}_\theta
            \right\} \ ,
\nonumber\\
F_{\theta\varphi}  & = & -\frac{n}{2}\left\{
         \left(2 \snt (H_2-1) +2 \ct H_3 -2\st H_2 (1-H_4) 
                                    +2 \st \partial_\theta H_3\right)
            \tau^{(2,n)}_r
            \right.
\nonumber\\ & & 
            \left.
        +\left(2 \cnt (H_2-1) +2 \ct (1-H_4) +2\st H_2 H_3
                                    -2 \st \partial_\theta H_4\right)
            \tau^{(2,n)}_\theta
            \right\} \ ,
\label{Fij}
\end{eqnarray}
and the covariant derivative of the Higgs field 
\begin{eqnarray}
D_r\Phi  & = & \frac{1}{r}\left\{
                   \left( r \partial_r \Phi_1+ H_1 \Phi_2 \right)  
                            \tau^{(2,n)}_r
                  +\left( r \partial_r \Phi_2- H_1 \Phi_1 \right)  
                            \tau^{(2,n)}_\theta
              \right\} \ ,
\nonumber\\
D_\theta\Phi  & = & \left(\partial_\theta \Phi_1 - 2 H_2 \Phi_2\right) 
                            \tau^{(2,n)}_r
                   +\left(\partial_\theta \Phi_2 + 2 H_2 \Phi_1\right)                             
                            \tau^{(2,n)}_\theta \ ,
\nonumber\\
D_\varphi\Phi  & = & n\left\{\left(\snt -2 \st (1-H_4)\right)\Phi_1
                   +\left(\cnt +2 \st H_3\right) \Phi_2\right\}
                            \tau^{(n)}_\varphi \ .
\label{coD}
\end{eqnarray}
The dimensionless energy density then becomes 
\begin{eqnarray}
\varepsilon & = & 
{\rm Tr} \left\{ 
\frac{1}{r^2}F_{r\theta}^2
+\frac{1}{r^2\sin^2\theta}F_{r\varphi}^2
+\frac{1}{r^4\sin^2\theta}F_{\theta\varphi}^2 
\right\}
\nonumber\\
& &
+\frac{1}{4}
{\rm Tr} \left\{
(D_r\Phi)^2
+ \frac{1}{r^2}(D_\theta\Phi)^2
+ \frac{1}{\sin^2\theta r^2}(D_\varphi\Phi)^2
\right\}
\nonumber\\
& &
-\frac{\kappa}{4}
{\rm Tr} \left\{ \frac{1}{r^2}[D_r\Phi, D_\theta \Phi]^2
       +\frac{1}{r^2\sin^2\theta}[D_r\Phi, D_\varphi \Phi]^2
       +\frac{1}{r^4\sin^2\theta}[D_\theta\Phi, D_\varphi \Phi]^2
         \right\}
+\lambda
(\left(|\Phi|^2-1\right)^2 \ ,
\label{energy}
\end{eqnarray}
where $|\Phi|=\sqrt{\Phi_1^2+\Phi_2^2}$ denotes the modulus of the Higgs field.

For a monopole-antimonopole pair we expect
a magnetic dipole field for the asymptotic gauge potential.
The dipole moment $C_{\rm \bf m}$ can be extracted from the gauge field
function $H_3$, in the gauge where the Higgs field approaches asymototically a
constant. Like in Ref.~\cite{KK} we find
\begin{equation}
H_3 = \frac{C_{\rm \bf m}}{r} \sin\theta
\ , \end{equation}
while all other gauge field functions decay faster.

\section{Numerical Results}

As noted in \cite{KK} the Ansatz Eqs.~(\ref{A_an}), (\ref{Phi_an})
possesses a residual $U(1)$ gauge symmetry.
To obtain an unique solution we use the
gauge fixing condition \cite{KK}
\begin{equation}
G_{\rm f} =
\frac{1}{r^2}\left( r\partial_r H_1 -2 \partial_\theta H_2\right) =0 \ .
\end{equation}
The system of partial differential equations is solved numerically
subject to the following boundary conditions,
which respect finite energy and finite 
energy density conditions as well as regularity and symmetry requirements.
These boundary conditions are at the origin
\begin{equation}
H_1(0,\theta)=H_3(0,\theta)=0\ , \ \ \ \ H_2(0,\theta)=H_4(0,\theta)=1 \ ,
\label{BCH0}
\end{equation}
\begin{equation}
\sin 2\theta \Phi_1(0,\theta)+\cos 2\theta \Phi_2(0,\theta) = 0 \ , \ \ \ \ 
\partial_r (\cos 2\theta \Phi_1(0,\theta)
 -\sin 2\theta \Phi_2(0,\theta)) = 0 \ ,
\label{BCP0}
\end{equation}
at infinity
\begin{equation}
H_1(\infty,\theta)=H_2(\infty,\theta)=0\ , \ \ \ \ 
H_3(\infty,\theta)=\sin\theta \ , \ \ 
\left(1-H_4(\infty,\theta)\right)=\cos\theta
\end{equation}
\begin{equation}
\Phi_1(\infty,\theta)=1\ , \ \ \ \ \Phi_2(\infty,\theta)=0 \ ,
\end{equation}
and on the $z$-axis
\begin{equation}
H_1(r,\theta=0,\pi) = H_3(r,\theta=0,\pi) 
 = \partial_\theta H_2(r,\theta=0,\pi)
 = \partial_\theta H_4(r,\theta=0,\pi) = 0 \ ,
\end{equation}
\begin{equation}
\Phi_2(r,\theta=0,\pi)
 = \partial_\theta \Phi_1(r,\theta=0,\pi)=0\ .   
\end{equation}

The numerical calculations were performed with the software package
CADSOL, based on the Newton-Raphson method \cite{schoen}. We have carried out
the main part of the numerical analysis for the case of unit vortex number
$n=1$ in (\ref{n}) as in Refs.~\cite{Rueb} and \cite{KK}. In addition we
have also studied more briefly, the case of $n=2$.

Starting with the case of vorticity $n=1$,
we have constructed monopole-antimonopole solutions for a large range
of values of the coupling constant $\kappa$. For vanishing coupling
constant $\kappa$ the monopole-antimonopole solution corresponds to a
non-Bogomol'nyi solution of the BPS system, for which our results are in good
agreement with those of \cite{KK}. Our numerical analysis was carried out for
the skyrmed model in the absence of the Higgs potential, namely with
$\lambda=0$ in (\ref{energy}). We did however check that the presence of
nonvanishing $\lambda$ does not change the qualitative properties of our
solutions. As expected the only effect it has is in the large $r$ asymptotic
region, where the modulus of the Higgs field for example, reaches its
asymptotic value faster, namely exponentially.

In Figure~1 we show the normalised energy of the solitons $E/4\pi\eta$
and the energy $E_{inf}/4\pi\eta$, of the monopole-antimonopole pair
with infinite separation corresponding to twice the energy of a
charge-1 monopole, as functions of the coupling constant $\kappa$. As
can be seen from Figure~1 the energy of the monopole-antimonopole
solution is less than the energy of a monopole-antimonopole pair with
infinite separation for all values of $\kappa$.
\begin{center}\mbox{
\epsfxsize=8.cm\epsffile{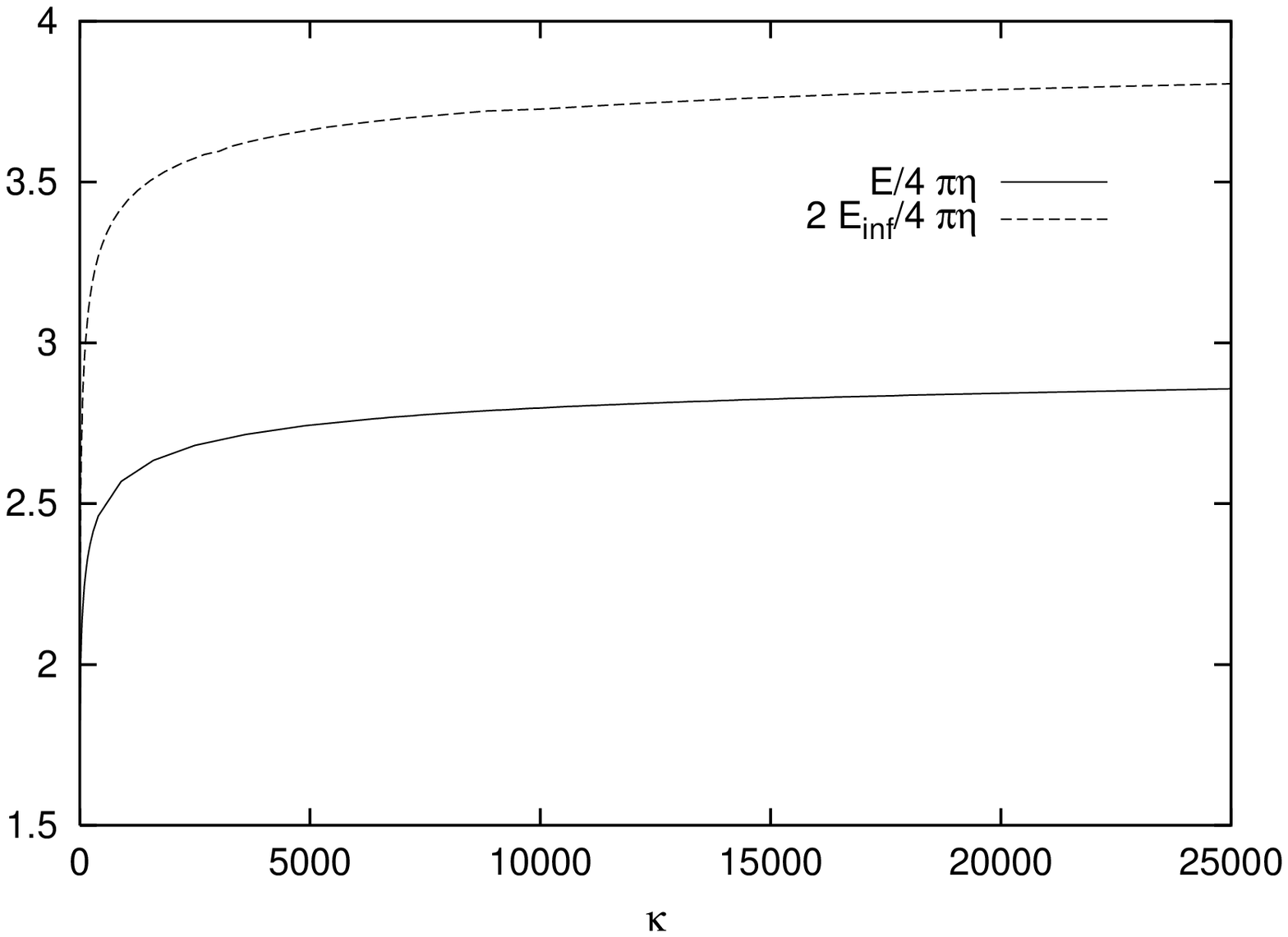}}\\
Figure~1: The energy of the monopole-antimonopole solution (solid line)
and the energy of a monopole-antimonopole pair with infinite
separation (dashed line), for $n=1$.
\end{center}
In Figure~2 we exhibit the modulus of the Higgs field $|\Phi(\rho,z)|$
as a function of the coordinates $\rho=\sqrt{x^2+y^2}$ and $z$ for $\kappa=0$ 
and $\kappa=100$. The zeros of $|\Phi(\rho,z)|$ are located on the
positive and negative $z$-axis at $\pm z_0 \approx 2.1 $ for $\kappa=0$ 
and at $\pm z_0 \approx 1.5 $ for $\kappa=100$. The distance $d$ of the
two zeros of the Higgs field decreases monotonically with increasing
$\kappa$.

\begin{center}\mbox{
\epsfxsize=8.cm\epsffile{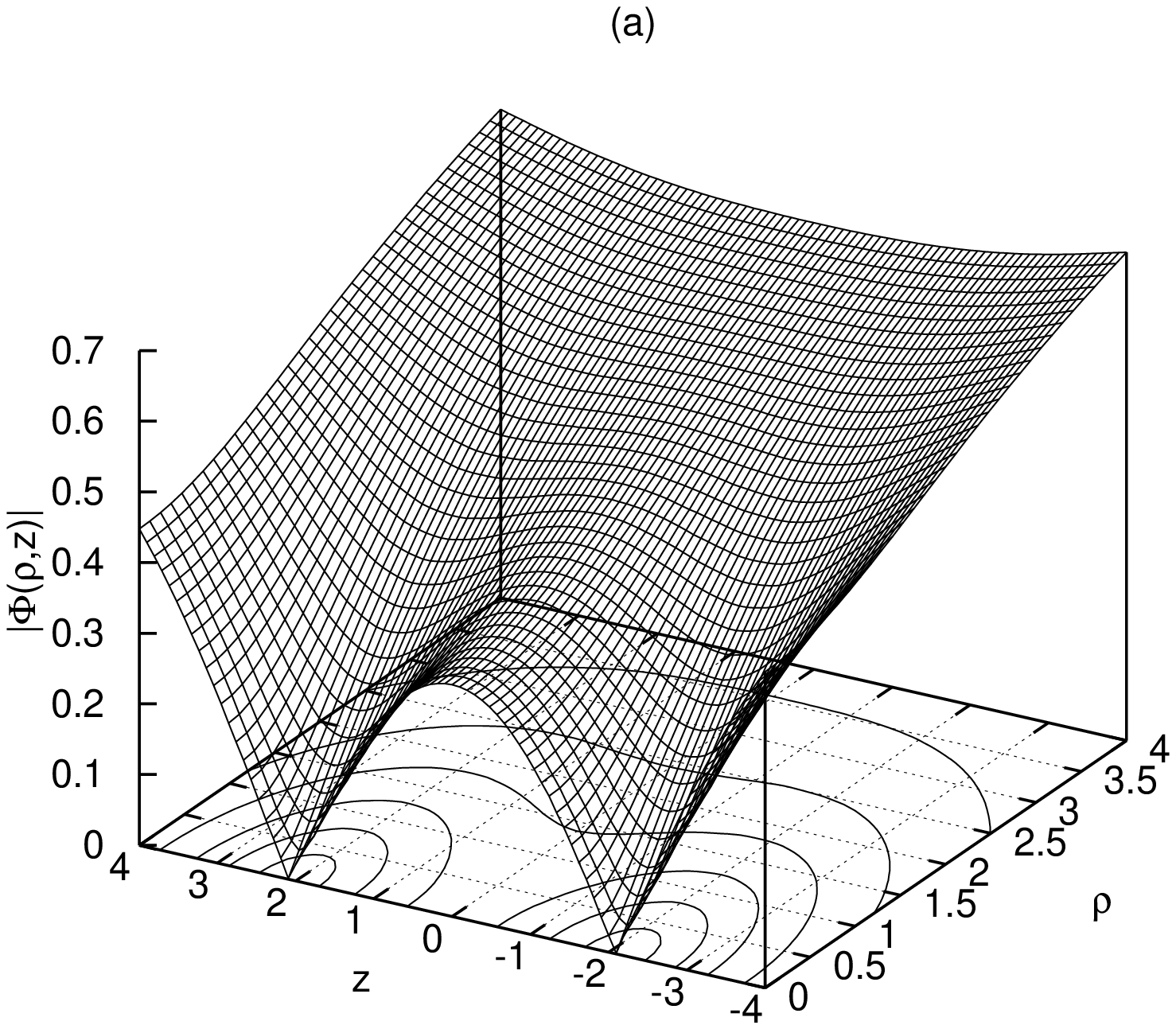}
\epsfxsize=8.cm\epsffile{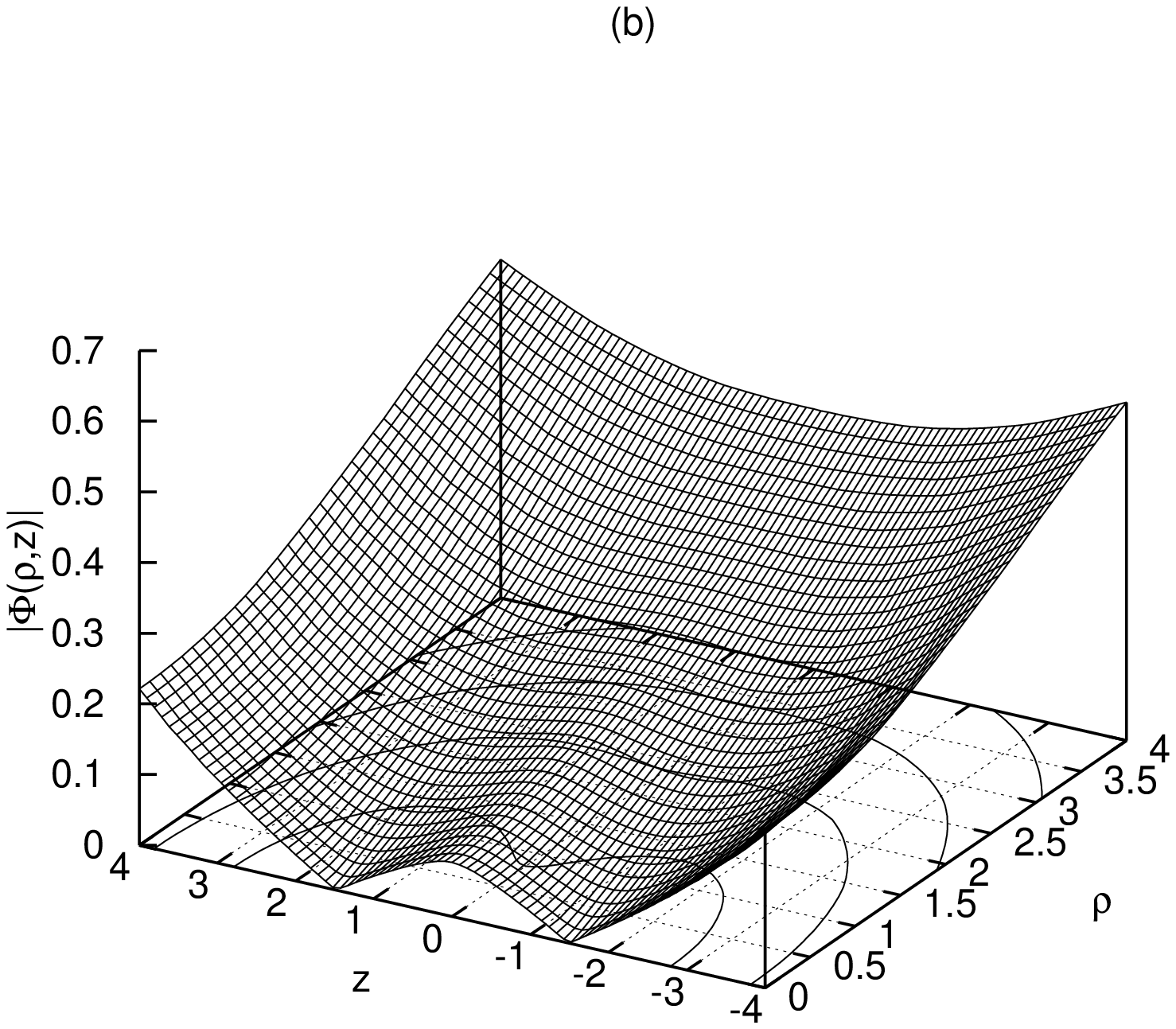}}\\
Figure~2: The modulus of the Higgs field as a function of $\rho$ 
and $z$ for $\kappa=0$ (a) and $\kappa=100$ (b), for $n=1$
\end{center}
Asymptotically $|\Phi(\rho,z)|$ approaches the value 1. But at the
origin  the value of the modulus of the Higgs field  decreases monotonically
with increasing $\kappa$ (see Figure~3). In the
limit $\kappa \rightarrow \infty$ $|\phi_0| \approx 0.015$, and we
expect the modulus of the Higgs field to be very small for $|z| \le 4$.

\begin{center}\mbox{
\epsfxsize=8.cm\epsffile{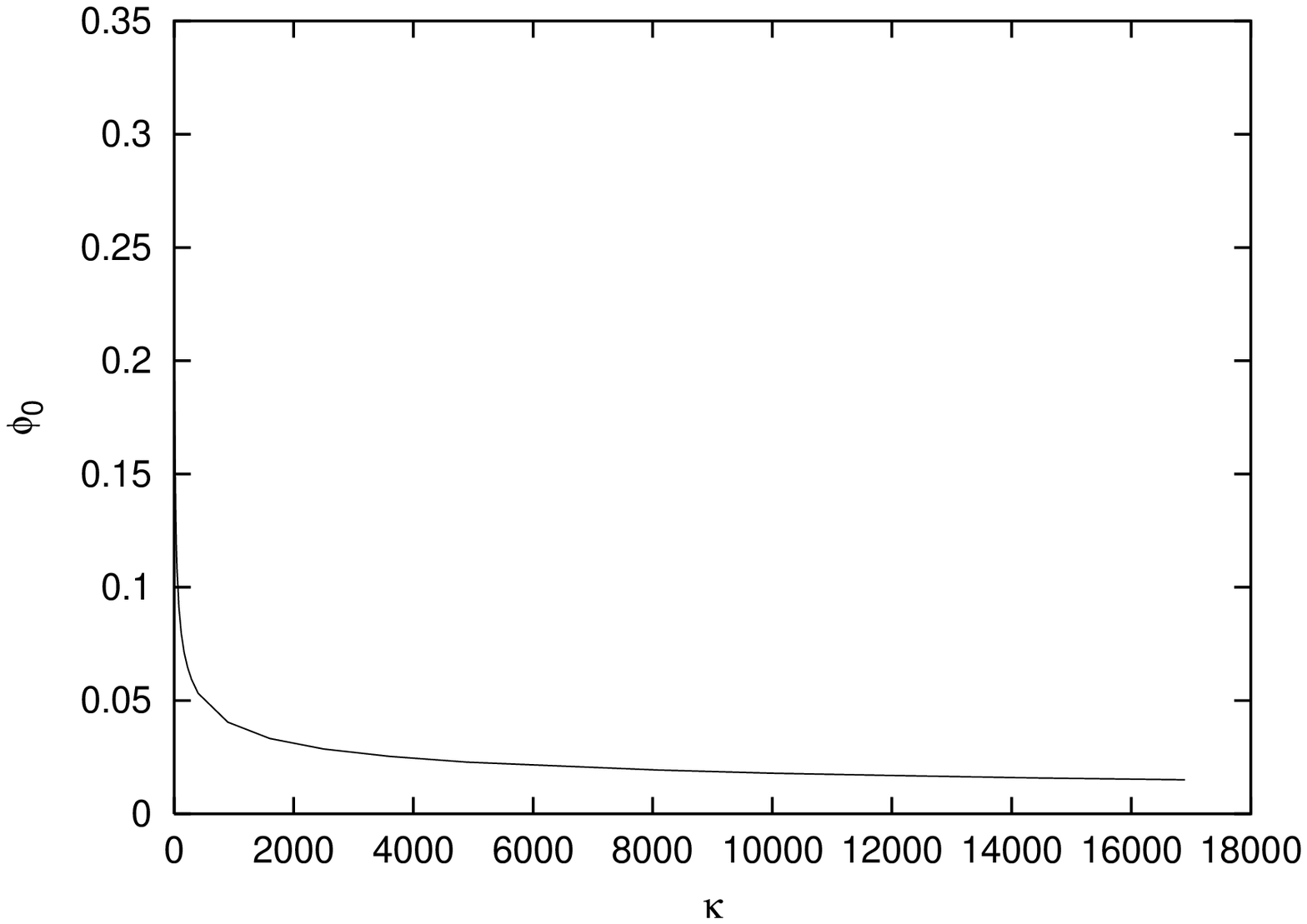}}\\
Figure~3: The modulus of Higgs fields at the origin as a function of $\kappa$,
for $n=1$
\end{center}
In Figure~4 we show the energy density of the monopole-antimonopole
solution as a function of the coordinates $\rho=\sqrt{x^2+y^2}$ and
$z$ for $\kappa=0$ and $\kappa=100$. At the locations of the Higgs
field the energy density posesses maxima.

For small values of coupling constant $\kappa$ the equal energy
density surfaces near the locations of the zeros of the Higgs field
assume a shape close to a sphere, centered at the location of the
respective zero (see Figure~4 (a)). This presents further support for the
conlcusion, that at the two zeros of the Higgs field a monopole and an
antimonopole are located, which can be clearly distinguished from each
other, and which together form a bound state.
\begin{center}\mbox{
\epsfxsize=8.cm\epsffile{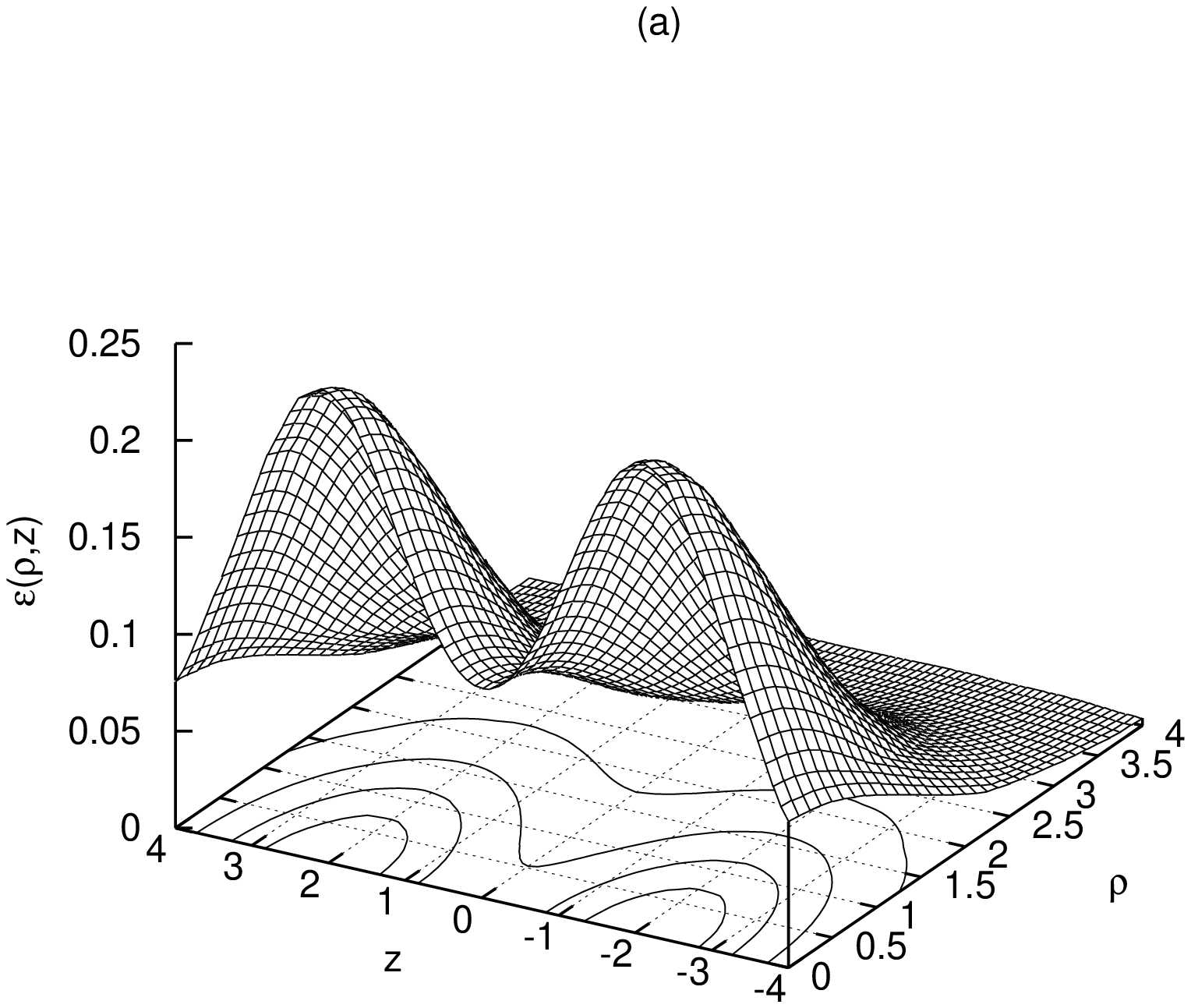}
\epsfxsize=8.cm\epsffile{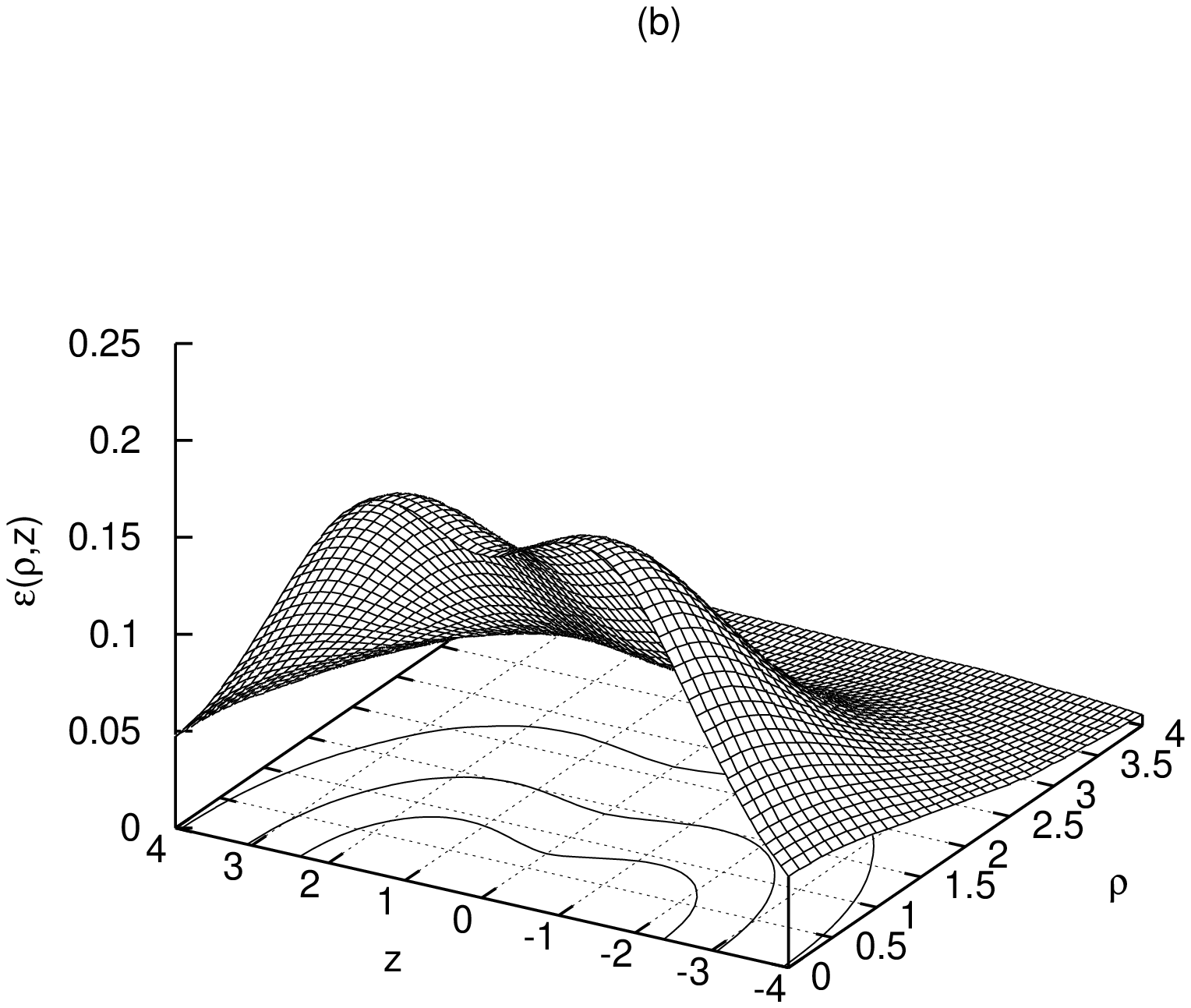}}\\
Figure~4: The dimensionless energy density as a function of $\rho$ 
and $z$ for $\kappa=0$ (a) and $\kappa=100$ (b), for $n=1$
\end{center}
With increasing $\kappa$ the distance $d$ between the monopole anti-monopole
centres becomes smaller tending to a limit as $\kappa\to\infty$. At the
same time the spherical equal energy surfaces in Figure~4(b) become larger,
and the equal energy
density surfaces assume a shape that looks like the intersection of two spheres
(see Figure~4 (b)), thus making it more difficult to distinguish the monopole
from the anti-monopole. The dependence of the separation length $d$ is given
in Table~1 below as a function of $\kappa$.

Having exibited the qualitative properties of our {\it dipole} solutions, we
give the values of the dipole moment that we calculated as a function of the
coupling constant $\kappa$, again in the Table~1. As expected, with decreasing
$d$ the dipole moment $C_{\rm \bf m}$ also decreases.
\begin{center}\mbox{
\begin{tabular}{|c|c c c c c c c c c c|}
 \hline
$\kappa$ & 0 & 9 & 16 & 25 & 36 & 49 & 64 & 100 & 8100 & 10000\\
\hline
  $d$  & 4.19 & 3.64 & 3.49 & 3.38 & 3.29 & 3.21 & 3.16 & 3.06 & 2.54 & 2.53    \\
\hline
$C_{\rm \bf m}$ & 2.36 & 2.27 & 2.23 & 2.19 & 2.15 & 2.11 & 2.07 & 2.02 & 1.66 & 1.65 \\
 \hline
\end{tabular}}\\

\medskip
Table~1: Monopole anti-monopole separation $d$ and
dipole moment $C_{\rm \bf m}$ as a functions of $\kappa$
\end{center}
Finally we constructed solutions for the case of vorticity $n=2$. Most of the
qualitative properties of these solutions do not differ from those of the
$n=1$ case just described. The most noticable quantitative difference concerns
the value of the modulus of the Higgs field at the origin, analogous with
Figures~1(a) and 1(b). We do not exhibit here these analogous figures, but
simply note that the the moduli of the Higgs fields at the origin are
{\it smaller} than those in Figures~1(a) and 1(b) for the same values of
the coupling constant $\kappa$.

Another difference, qualitative though expected, is that the surfaces of
equal energy are not spheres centred on the $z$-axis but describe rings or tori
around it. This is exhibited in Figures~5(a) and 5(b), analogously with
Figures~4(a) and 4(b).

\begin{center}\mbox{
\epsfxsize=8.cm\epsffile{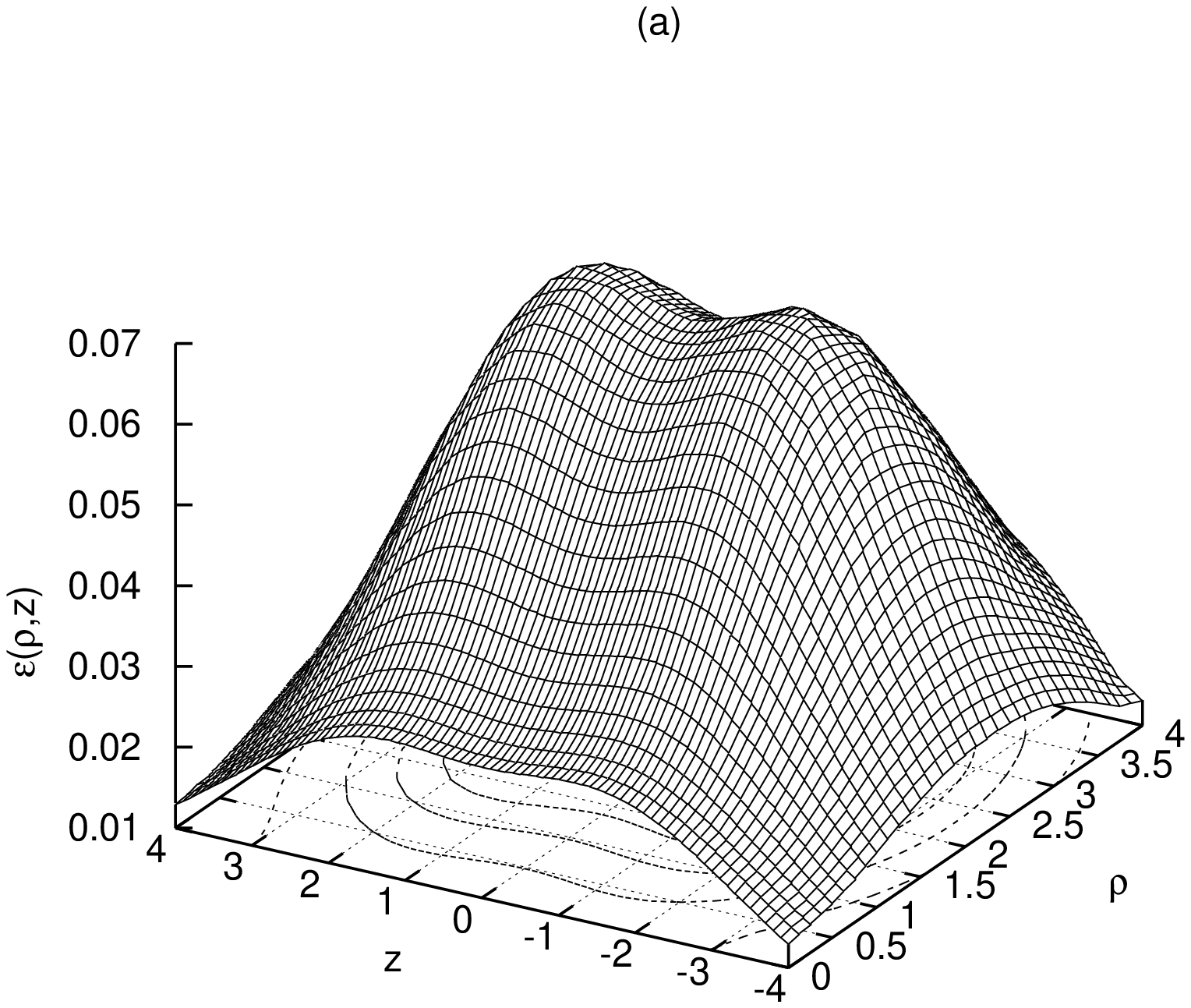}
\epsfxsize=8.cm\epsffile{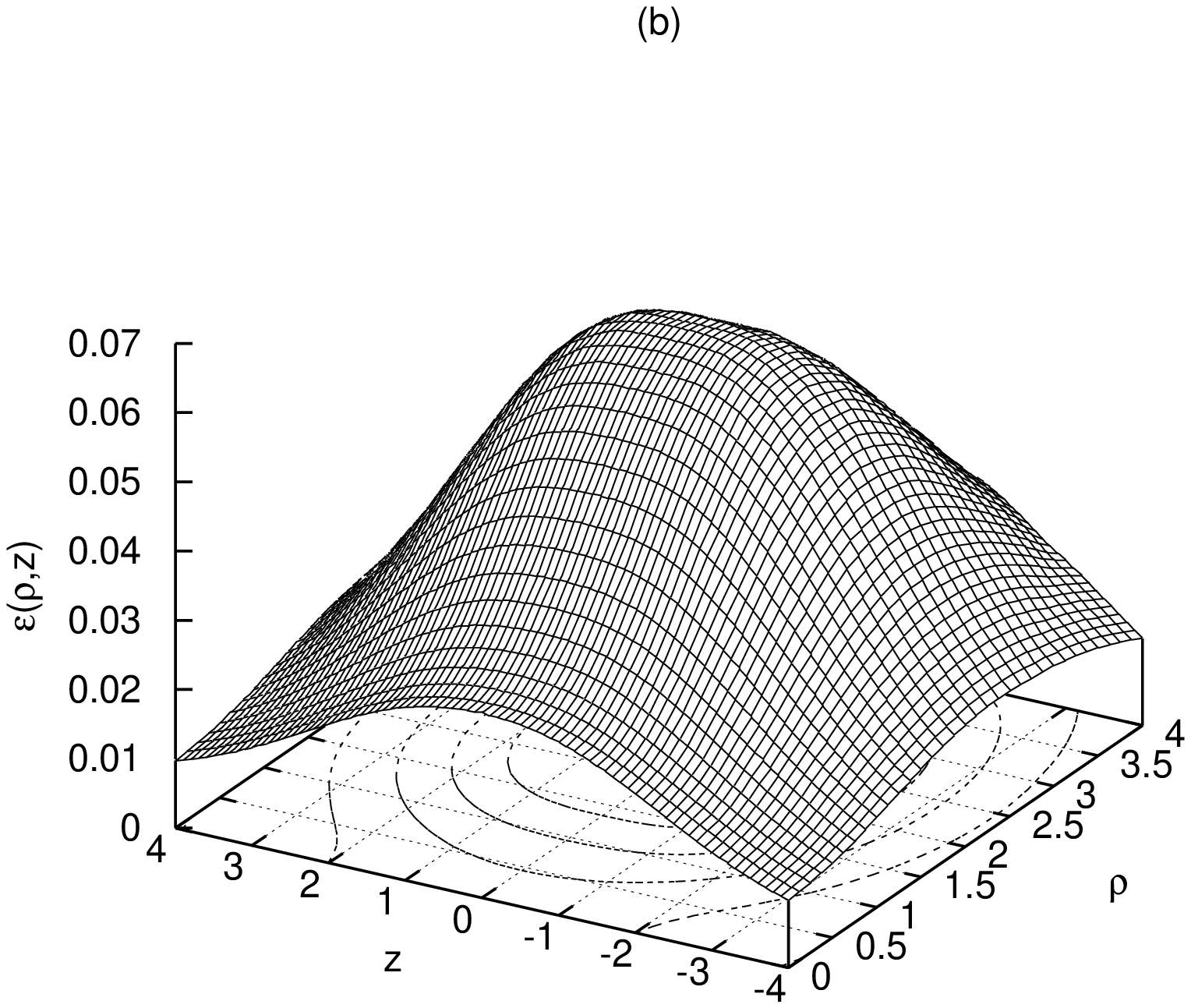}}\\
Figure~5: The dimensionless energy density as a function of $\rho$ 
and $z$ for $\kappa=0$ (a) and $\kappa=100$ (b), for $n=2$
\end{center}
Again, as $\kappa$ grows, the distiction between the monopole and anti-monopole
rings gets blurred.

\section{Summary}
We have contructed axially symmetric solutions to a simple $SU(2)$ skyrmed
YM-Higgs model, with such boundary conditions that result in the description
of a monopole anti-monopole pair with zero magnetic charge. These solutions
have lower mass than two infinitely separated charge-1 monopoles, and since
they are characterised by zero magnetic charge, are not topologically stable.

When the usual boundary conditions are imposed, the skyrmed $SU(2)$
YM-Higgs model employed here supports mutually attractive monopoles, including
axially symmetric charge-2 monopoles. This is in contrast to the Georgi-Glashow
model studied in \cite{KK} where due to the Higgs potential the monopoles
are mutually repulsive~\cite{KKT}. Nevertheless, the qualitative features
of the monopole anti-monopole solutions in the two models are similar.
Increasing the Skyrme coupling constant $\kappa$ in the present model results
in the approaching of the monopole and the anti-monopole centres down to a
limiting value 2.53 as $\kappa\to\infty$, just as it does to the limiting 
value 3.0 as $\lambda\to\infty$ in the Georgi-Glashow model
$\lambda$ being the Higgs coupling constant. (Our results are for $\lambda=0$.)

Another parallel property in the two models is the changing dipole moment
with respect to the change in the Skyrme coupling constant $\kappa$ and
the Higgs coupling constant $\lambda$, in the two models respectively.
Specifically in the present model the magnetic moment decreases with increasing
$\kappa$, with limiting value 1.64, while in the Georgi-Glashow model it
decreases with increasing $\lambda$, with limiting value 1.55,
in the same units.

Finally, we studied also the case of a zero charge monopole which has
vortex number $n=2$ rather than $n=1$. The qualitative
properties again stay unchanged. The most noticable quantitative difference
of the $n=2$ soltion is that the value of the modulus of the Higgs field at
the origin is smaller than that of the $n=1$ solution, for the same value of
$\kappa$, and, the distance between the two centres is also smaller. For
example at $\kappa=25$ the distance $d=3.38$ for the $n=1$ solutions while
that for the $n=2$ is $d=1.33$.

\vspace{1cm}

{\large \bf Acknowledgments}

It is a pleasure for us to thank Burkhard Kleihaus for numerous valuable
discussions and for his generous help. This work was carried out in the
framework of Enterprise--Ireland project SC/2000/020 and IC/2002/005..

\small{

\end{document}